\documentclass[12pt]{article2}
\usepackage{citesupernumber}
\usepackage[T1]{fontenc}
\usepackage[utf8]{inputenc}
\usepackage{authblk}
\usepackage{setspace}
\usepackage{fullpage,amssymb,amsmath}
\usepackage{caption}
\usepackage{subfigure, epsfig}
\usepackage{pdfpages}
\title{Planetary System Disruption by Galactic Perturbations to Wide Binary Stars} 
\author[1,2,3]{Nathan A. Kaib}
\author[4,5]{Sean N. Raymond}
\author[1]{Martin Duncan}
\affil[1]{Department of Physics, Queen's University, Kingston, ON K7L 3N6, Canada}
\affil[2]{Canadian Institute for Theoretical Astrophysics, University of Toronto, Toronto, ON M5S 3H8, Canada}
\affil[3]{Dept of Physics and Astronomy \& Center for Interdisciplinary Exploration and Research in Astrophysics (CIERA), Northwestern University, Evanston, IL 60208, USA}
\affil[4]{Univ. Bordeaux, LAB, UMR 5804, F-33270, Foirac, France}
\affil[5]{CNRS, LAB, UMR 5804, F-33270, Floirac, France}
\date{}

\begin{document} 

\maketitle

\doublespacing

{\bf Nearly half of the exoplanets found within binary star systems reside in very wide binaries with average stellar separations beyond 1,000 AU\cite{roe12} (1 AU being the Earth-Sun distance), yet the influence of such distant binary companions on planetary evolution remains largely unstudied. Unlike their tighter counterparts, the stellar orbits of wide binaries continually change under the influence of the Galactic tide and impulses from other passing stars. Here we report numerical simulations demonstrating that the variable nature of wide binary star orbits dramatically reshapes the planetary systems they host, typically Gyrs after formation. Contrary to previous understanding, wide binary companions may often strongly perturb planetary systems, triggering planetary ejections and exciting orbital eccentricities of surviving planets. Indeed, observed exoplanet eccentricities offer evidence of this; giant exoplanet orbits within wide binaries are statistically more eccentric than those around isolated stars. Both eccentricity distributions are well-reproduced when we assume isolated stars and wide binaries host similar planetary systems whose outermost giant planets are scattered beyond $\sim$10 AU from their parent stars via early internal instabilities.  Consequently, our results suggest that although wide binaries eventually truncate their planetary systems, most isolated giant exoplanet systems harbor additional distant, still undetected planets.}

Unlike binaries with separations below $\sim$10$^3$ AU, very wide binary stars are only weakly bound by self-gravity, leaving them susceptible to outside perturbations. As a result, the Milky Way's tide and impulses from other passing stars strongly perturb wide binary  orbits \cite{hegras96, jiatre10}. These perturbations, which are fairly independent of the orbiting object's mass, are also known to dramatically affect the dynamics of solar system comets with similar orbital distances\cite{oort50,heitre86}.  Galactic perturbations drive a psuedo-random walk in these comets' pericenters (or closest approach distances to the central body)\cite{oort50,kaibquinn09}.  The same will occur in wide binary orbits. Thus, even if a very wide binary's initial pericenter is quite large, it will inevitably become very small at some point if it remains gravitationally bound and evolves long enough. Such low pericenter phases will produce close stellar passages between binary members, with potentially devastating consequences for planetary systems in these binaries\cite{adamslaugh01,zaktre04}. Counterintuitively, we therefore suspect that wide binary companions could more dramatically affect planetary system evolution than tight binaries.

To investigate this scenario, we use the MERCURY simulation package to perform 2600 simulations modeling the orbital evolution of our Sun's four giant planets (on their current orbits) in the presence of a very wide binary companion\cite{cham02}. These simulations are listed as set ``A'' in Table 1, which briefly summarizes our different simulations' initial conditions (supplementary information provides details).  An example simulation is shown in Figure 1.  Initially, the binary companion has no effect on the planets' dynamics since its starting pericenter ($q$) is $\sim$3,000 AU.  However, after 1 Gyr of evolution, galactic perturbations drive the binary pericenter near 100 AU, exciting the eccentricities of Neptune and Uranus.  Once again at 3.5 Gyrs, the binary passes through another low pericenter phase, this time triggering the ejection of Uranus.  Finally at 7.2 Gyrs, the binary makes a final excursion to low $q$, causing Neptune's ejection.

Such behavior is not unusual.  Depending on the binary's mass and semimajor axis (mean separation, or $a_{*}$), Figure 2a demonstrates that $\sim$30--60\% of planetary systems in simulation set A experience instabilities causing one or more planetary ejections after 10 Gyrs (the approximate age of our galaxy's thin disk).  Even though binaries with smaller semimajor axes are less affected by galactic perturbations, Figure 2a shows the influence of binary semimajor axis on planetary instability rates is weak.  This is because tighter binaries make pericenter passages at a higher frequency.  In addition, when they reach low-$q$ phases they remain stuck there for a much longer time than wider binaries.  As Figure 2b shows, both of these effects cause tighter binaries to become lethal at a much larger pericenter, offsetting the Galaxy's diminished influence.  The large majority of binary-triggered instabilities are very delayed.  For binaries with $a_{*}\gtrsim2,000$ AU, Figure 2c shows that well over 90\% of instabilities occur after at least 100 Myrs of evolution, well after planet formation is complete.  For tighter binaries, many more begin in orbits that destabilize the planets nearly instantly.  

While planets are believed to form on nearly circular orbits\cite{liss93}, most known giant planets ($m \sin{i} > 1$ M$_{\rm Jup}$) have significant non-zero orbital eccentricities (eccentricities of less massive planets are known to be colder)\cite{wright09}. This observed eccentricity distribution can be reproduced remarkably well when systems of circularly orbiting planets undergo internal dynamical instabilities causing planet-planet scattering events that eject some planets and excite the survivors' eccentricities\cite{jurtre08,fordras08,malmdav09,ray10}.  For planetary systems within wide binaries, Figure 2a predicts that many should undergo additional dynamical instabilities triggered by their stellar companions.  Thus, these systems should experience an even greater number of planet-planet scattering events than isolated planetary systems. 

This raises the possibility that the eccentricities of exoplanets may hold a signature of the dynamical process illustrated in Figure 1.  Indeed, the overall distribution of exoplanet eccentricities provides compelling evidence of our disruptive mechanism.  Figure 3a compares the observed eccentricity distribution of all Jovian-mass ($m \sin{i} > 1$ M$_{\rm Jup}$) exoplanets found in binaries\cite{roe12} with the distribution of Jovian-mass planets around isolated stars.    As can be seen, the distribution of planets within wide binaries is significantly hotter than planetary systems without known stellar companions. A Komolgorov-Smirnov test returns a probability (or $p$-value) of only 0.6\% that such a poor match between the two datasets will occur if they sample the same underlying distribution. Thus, we reject the null hypothesis that the distributions are the same.  Although it consists of just 20 planets, our wide binary planetary sample contains the two most eccentric known exoplanet orbits, HD 80606b and HD 20782b (see Figure 2). Furthermore, these excited eccentricities seem to be confined to only very wide binary systems.  Figure 3a also shows the eccentricity distribution of planets residing in binaries with average separations below $10^3$ AU.  Unlike wider binaries, here we see that these eccentricities match very closely with the isolated distribution.  (A K-S test returns a $p$-value of 91\%.)  This suggests the variable nature of distant binary orbits is crucial to exciting planetary orbits.  Large eccentricities of planets within binaries have previously been explained with the Kozai resonance\cite{koz62,hol97,wumur03,fabtre07}, yet this effect should be most evident in these tighter binary systems.  

We perform additional simulations attempting to explain the observed eccentricity excitation in Figure 3a with the mechanism illustrated in Figure 1.  These additional simulation sets are summarized in Table 1 (B1--B3).  Unlike the internally stable planetary systems in simulation set A, these simulated systems consist of 3 approximately Jovian-mass planets started in unstable configurations (to induce planet-planet scattering) and evolved for 10 Gyrs (see supplementary information).  In the simulation sets presented in Figure 3b, we naturally reproduce both observed eccentricity distributions using {\it the same initial planetary systems}.  When our planetary systems are run in isolation (set B1 in Table 1) planet-planet scattering caused by internal instabilities yields the observed planetary eccentricities for isolated stars (K-S test $p$-value of 0.42).  Then when a 0.4 M$_{\sun}$ binary companion is added to each system (set B2 in Table 1) the eccentricity distribution is heated further, and again the match to observations is quite good, with a K-S test $p$-value of 0.46.  

In Figure 3c, the match to observed planetary eccentricities is much poorer.  Here we rerun our binary simulations with galactic perturbations shut off to yield static binary orbits (set B3 in Table 1).  In this case, the eccentricity distribution is barely more excited than the isolated cases, indicating that the variable nature of wide binary orbits is crucial to heating planetary eccentricities.  Otherwise, most stellar companions always remain far from the planets.  

In Figure 3d we reexamine simulation set B2 to determine which types of planetary systems are most influenced by wide binary companions.  By examining the planetary systems after only 10 Myrs, we can view them after most have experienced internal instabilities but before the binary has played a large role (since its effects are delayed).  We find that 70\% of our planetary systems have collapsed to two planets.  (The remaining are comprised of nearly equal numbers of 1- and 3-planet systems.)  We then split these two-planet systems into those with the outer planet beyond 10 AU and those with all planets confined inside 10 AU.  In Figure 3d, the final ($t=10$ Gyrs) eccentricity distribution is shown for both subgroups of planetary systems.  As can be seen, the more extended planetary systems eventually yield much more excited eccentricities compared to the compact systems.  This is because binaries do not have to evolve to such low pericenters to disrupt extended systems.  In fact, the observed wide binary planetary eccentricity distribution cannot be matched without using wide binaries with planets beyond 10 AU ($p = 0.016$ from a K-S test).  Assuming planets form similarly in wide binaries and isolated systems, the planetary eccentricity excitation observed within wide binaries may offer new constraints on the bulk properties of isolated giant exoplanet systems, which dominate the giant exoplanet catalog.  While most detection efforts are currently insensitive to planets with periods beyond $\sim$10 years, our work argues that massive longer period planets (beyond $\sim$10 AU) should be common around isolated stars.  Indeed, such distant planets have recently been directly observed\cite{mar10} and microlensing results suggest many such planets reside far from host stars\cite{sum11}. 

Due to the variable nature of their orbits, very distant binary companions may affect planetary evolution at least as strongly as their tighter counterparts.  This represents a paradigm shift in our understanding of planet-hosting binaries, since previous works tend to assume only tighter binaries strongly influence planetary system evolution\cite{egg07,desbar07}.  Intriguingly, the eccentricities of planets in wide binaries may provide new constraints on the intrinsic architectures of all planetary systems.  To further develop this prospect, searches for common proper motion companions to planet-hosting stars should be continued and expanded\cite{rag06,mug06,desbar07,egg07}.

\section*{Acknowledgements}
We thank John Chambers and Rok Ro{\v s}kar for discussions.  This work was funded by a CITA National Fellowship and Canada's NSERC.  SNR thanks the CNRS's PNP program and the NASA Astrobiology Institute's Virtual Planetary Laboratory team.  Our computing was performed on the SciNet General Purpose Cluster at the University of Toronto.

\section*{Author Contributions}
NAK performed the simulations and analysis and was the primary writer of this paper. SNR and MJD helped initiate the project and advised on simulations and analysis.

\clearpage

\begin{table}
\centering
\begin{tabular}{c c c c c c c}
\hline
Name & Number & Planet & Planet & Binary & Binary & External \\
 & of Planets & Masses & $a$-range & Mass & $a_{*}$ & Perturbations \\
 & &  (M$_{\rm Jup}$) & (AU) & (M$_{\sun}$) & (AU) & Included \\
\hline
A & 4 & SS & SS & 0.1 -- 1.0 & 1,000 -- 30,000 & Tide + Stars\\
B1 & 3 & 0.5 -- $\sim$15 & 2 -- $\sim$15 & None & None & None\\
B2 & 3 & 0.5 -- $\sim$15 & 2 -- $\sim$15 & 0.4 & 1,000 -- 30,000 & Tide + Stars\\
B3 & 3 & 0.5 -- $\sim$15 & 2 -- $\sim$15 & 0.4 & 1,000 -- 30,000 & None\\
\end{tabular}
\caption{{\bf - Initial Conditions of Simulation Sets.} SS refers to planetary systems resembling the solar system's four giant planets, and $a$ represents semimajor axis.  ``Tide + Stars'' refers to perturbations from the Galactic tide and passing field stars.}
\end{table}

\clearpage

\begin{figure}[p]
\centering
\includegraphics[scale=1.7]{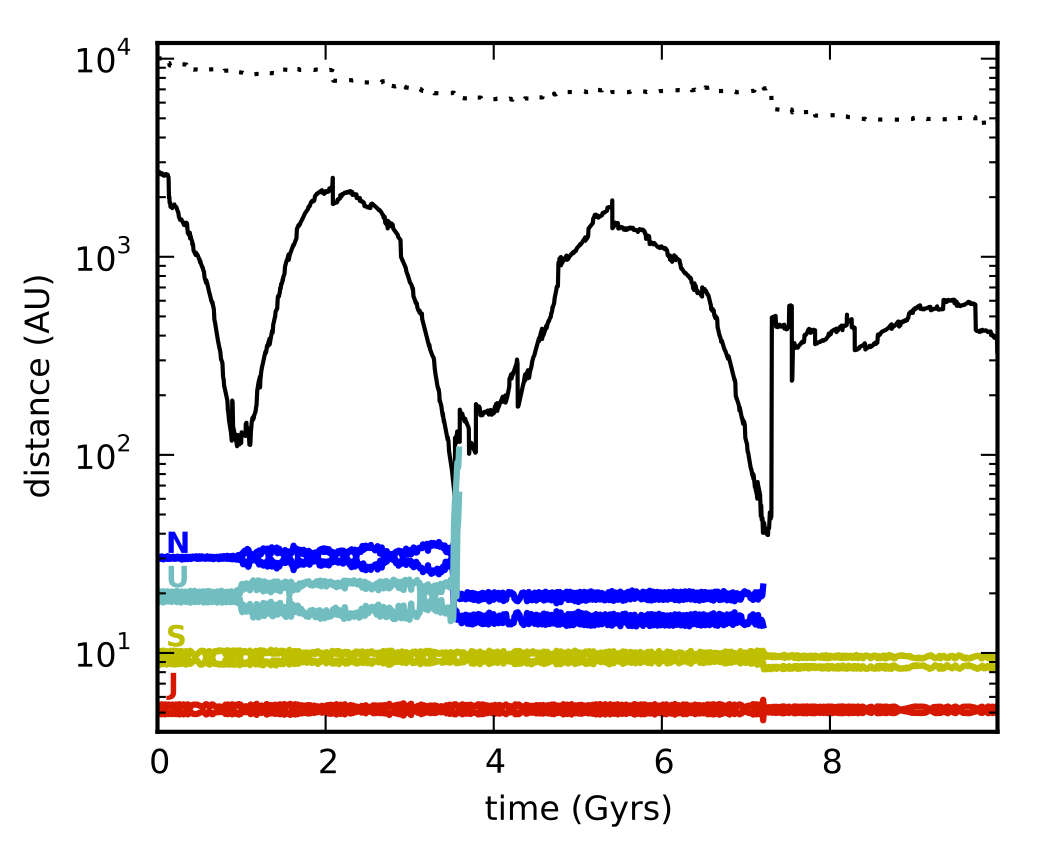}
\caption{Simulation of a binary-triggered instability in our solar system.  The pericenter and apocenter are plotted for Jupiter (red), Saturn (gold), Uranus (cyan), Neptune (blue).  The binary's semimajor axis (dotted black line) and pericenter (solid black line) are also shown.}\label{fig:1}
\end{figure}

\begin{figure}[p]
\centering
\includegraphics[scale=.65]{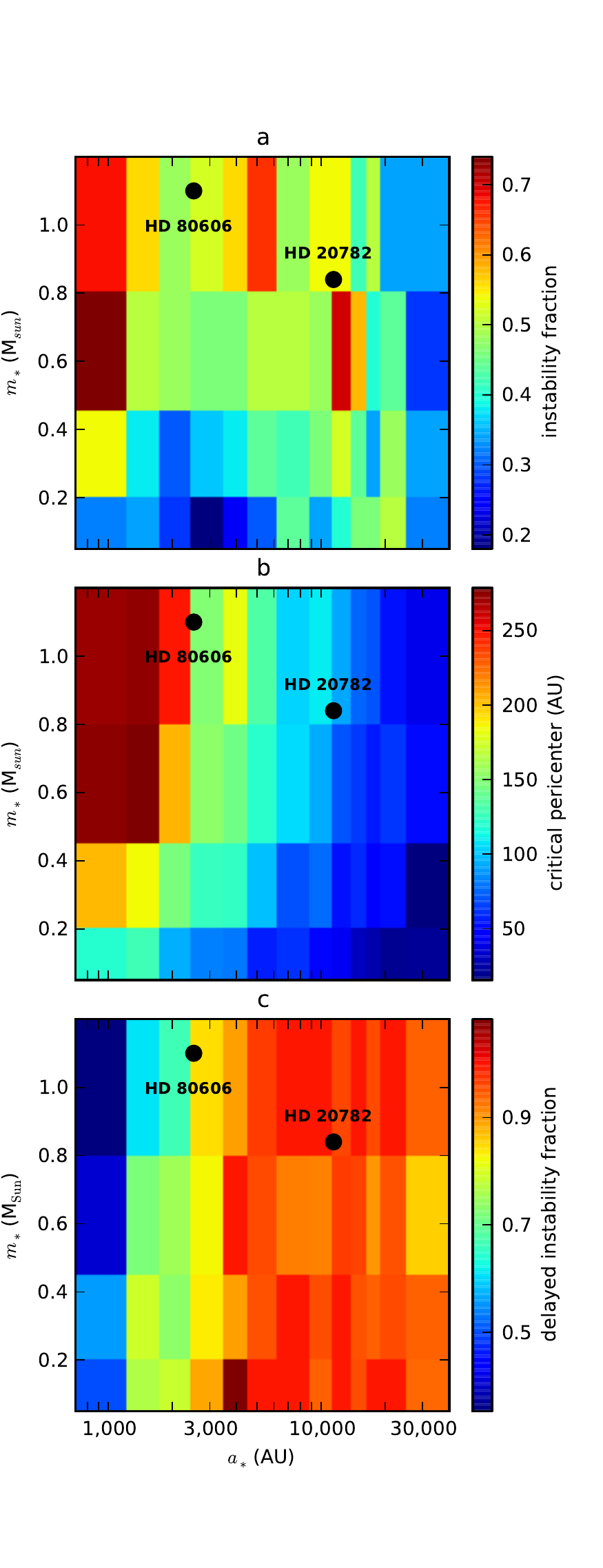}
\caption{{\footnotesize {\bf Disruption as a function of binary mass and separation.} {\bf a.} Map of the fraction of systems in set A (see Table 1) that lost at least one planet via instability. Binary mass is plotted on the $y$-axis, while the $x$-axis marks binary semimajor axis.  {\bf b.} The median binary pericenter below which an instability is induced in planetary systems as a function of binary mass and semimajor axis. {\bf c.} The fraction of instabilities that occur after the first 100 Myrs of evolution as a function of binary mass and semimajor axis. In each panel black data points mark the masses and presumptive semimajor axes of the HD 80606 and HD 20782 binaries, which host the two most eccentric known planetary orbits\cite{jones06,naef01}. While HD 80606b has been reproduced with a Kozai-driven mechanism, this process is markedly slower in even wider binaries such as HD 20782b\cite{inn97}. Moreover, the presence of more than one planet suppresses these Kozai oscillations\cite{inn97,bat11,kaib11}. However, our disruptive mechanism naturally collapses many systems to one planet, still enabling Kozai resonances to contribute to eccentricity excitation.  Panel {\bf a} of this figure suggests that binary-triggered instability rates become extremely high as binary semimajor axes drop below $\sim$10$^3$ AU, which could mean that tighter binaries trigger planetary system instabilities even more efficiently than those plotted here.  However, the initial conditions assumed for both our planetary orbits (solar system-like) and binary orbits (isotropic) become questionable for binary semimajor axes below $\sim$10$^3$ AU (see supplementary information).  Another interesting aspect not immediately obvious in panel {\bf c} is that instability times decreases at the largest binary semimajor axes.  This is because such binaries are rapidly unbound (or ``ionized'') by stellar impulses, making it impossible for these binaries to trigger instabilities at very late epochs (see supplementary information).}}
\end{figure}

\begin{figure}[p]
\centering
\includegraphics[scale=.65]{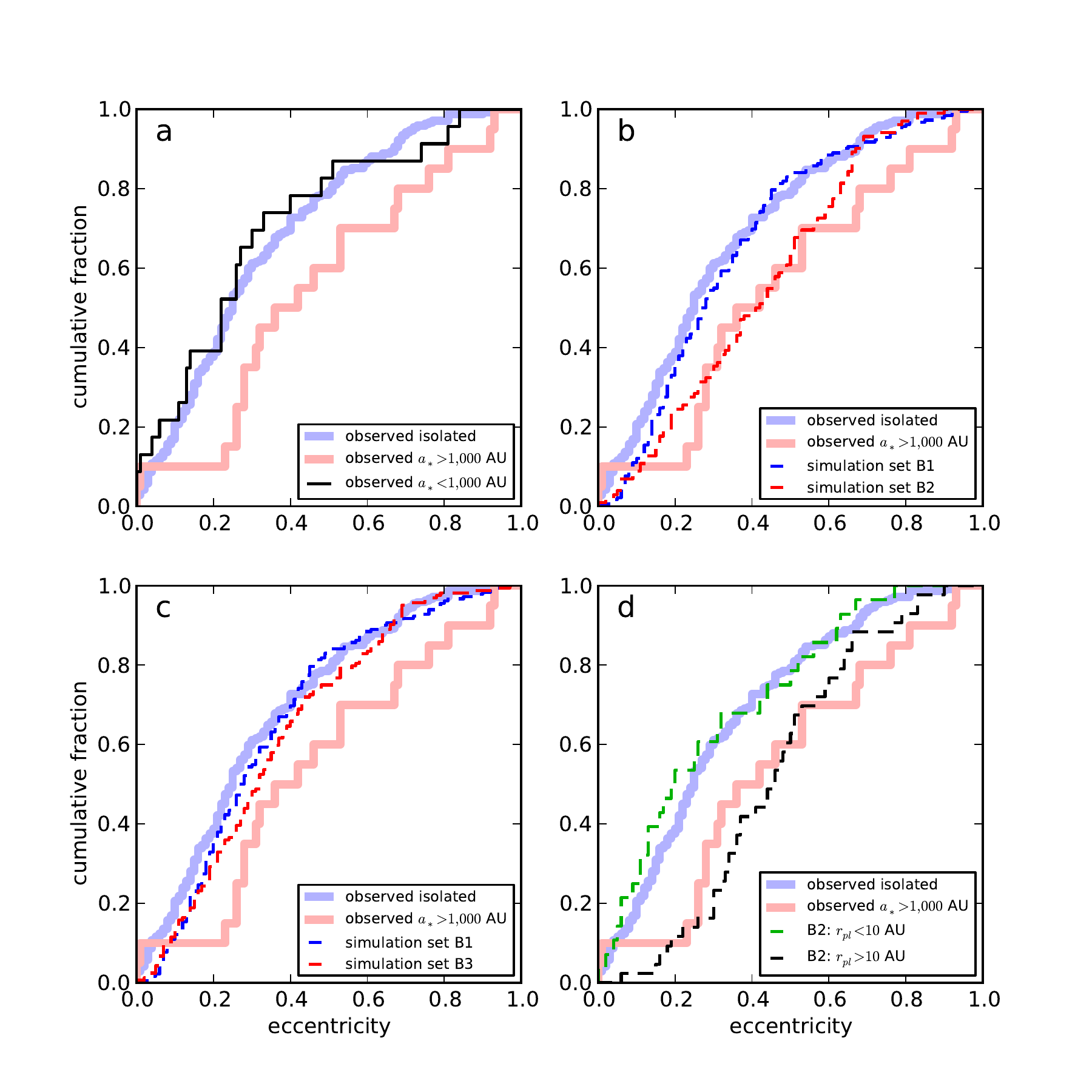}
\caption{{\bf Eccentricity excitation of planets of wide binaries.} Comparison of planetary eccentricity distributions.  Red lines correspond to systems with very wide ($a_{*}>10^3$ AU) companions and blue are isolated systems.  Additionally, solid lines mark observed distributions, while dashed lines describe the innermost planets in our simulated systems (sets B1--B3 in Table 1).  In all distributions, planets with $a<0.1$ AU are excluded to remove tidally circularized orbits. {\bf a.}  A comparison of observed exoplanet eccentricities within tighter ($a_{*}<10^3$ AU) binaries (black line) to those observed in very wide ($a_{*}>10^3$ AU) binaries and isolated systems.  {\bf b.} Eccentricities of simulated 3-planet systems after 10 Gyrs of evolution (sets B1 and B2 in Table 1). {\bf c.}  Simulations from panel B are rerun with no galactic perturbations (set B3 in Table 1). {\bf d.} The final eccentricities of two different subgroups of panel b's wide binary simulations:  systems that consisted of 2 planets extending beyond 10 AU at 10 Myrs (black), and 2-planet systems confined inside 10 AU at 10 Myrs (green).  One notices in panel {\bf b} that the presence of a wide binary does not seem to enhance the production of very extreme planetary eccentricities.  However, 1/4 of our wide binary systems have planets driven into the central star (1.7 times the rate within isolated systems).  Tidal dissipation not included in our models could strand these planets in very eccentric orbits before they collide with the central star\cite{wumur03,fabtre07}.  Interestingly, binaries also completely strip 20\% of our systems of planets, yielding naked stars that once hosted planets (see supplementary information).}
\end{figure}
\clearpage



\section*{Supplementary material}

# Supplementary Information

## 1 Outline

Here we present additional details on the work discussed in our main paper. This supplementary information is organized in the following sections:

- **Section 2 - Simulations Description**: Details of the numerical algorithms used to perform our simulations. This discussion includes our assumed orbital distribution of wide binaries and giant planets as well as our assumed initial planetary semimajor axes and spacing.

- **Section 3 - Supplementary Discussion**: Several abbreviated discussions from the main text concerning key results are expounded upon here. These include:

    - **3.1**: A justification of our choice of initial planetary orbits
    - **3.2**: How the initial distribution of planetary semimajor axes evolves over time.
    - **3.3**: Our choice of simulation integration times and how our results vary with shorter integrations.
    - **3.4**: Discussion of biases and contamination in the observed exoplanet distributions used in the main paper
    - **3.5**: A description of the evolution of our isolated planetary systems and their rate of internal instabilities
    - **3.6**: A systematic study of how binary-triggered instability rates vary with the radial distribution of planets
    - **3.7**: The fates of planets lost from our simulations
    - **3.8**: Stars that are completely stripped of planets by wide binary companions
    - **3.9**: The rate that wide binaries are ionized, or broken up by stellar perturbations
    - **3.10**: The potential effects of a more realistic, time-varying galactic environment
    - **3.11**: Exploratory simulations and a discussion of planetary system evolution within tighter binaries
    - **3.12**: Discussion of whether the excited planetary eccentricities in wide binaries could instead be explained by the wide binary star formation mechanism

- **Section 4 - Supplementary Table**: A list of the observed exoplanet systems in wide binaries (and their key characteristics) that were used to construct the observed eccentricity distributions of planets within binaries in the main paper.

## 2 Simulations Description

We have run two distinct types of simulations: those modeling the effects of a wide binary companion on planetary systems like our own solar system (set A in the main paper) and those modeling a wide binary's effects on generically generated planetary systems (sets B1–B3 in the main paper). These two batches are described in the following two subsections.

### 2.1 Solar System Simulations

For our simulations modeling the evolution of our solar system's four giant planets in the presence of a wide binary, we used the wide binary integrator included in the MERCURY simulation package[31, 32]. The giant planets were initially placed on their current orbits around the Sun in each simulation, while the binary companion's mass and initial orbit was varied. Four different binary companion masses were modeled: 0.1, 0.3, 0.6 and 1.0 $M_\odot$. In addition, thirteen distinct initial semimajor axes were used for the binary orbit: 1,000, 1,500, 2,000, 3,000, 4,000, 5,000, 7,500, 10,000, 12,500, 15,000, 17,500, 20,000, and 30,000 AU. All other initial binary

orbital elements were drawn from an isotropic distribution. For each mass-semimajor axis combination, 50 different simulations were run, varying the initial binary eccentricity, inclination, argument of pericenter and longitude of ascending node. This yielded a total of 2600 runs. Each simulation was integrated for 10 Gyrs using a 200-day timestep. Even though the age of the solar system is only 4.6 Gyrs, the Milky Way thin disk is thought to be ∼10 Gyrs old[33, 34].

To model galactic perturbations on the orbits of binaries, we use conditions that mimic the present day solar neighborhood. Although binaries near the Sun may have explored a large range of galactic environments[35], we choose to hold the galactic environments fixed to limit the number of parameters we explore. For the Galactic tide, we assume a local disk density of 0.1 $M_\odot/pc^3$, and our tidal model contains both a vertical and radial component[36]. However, the vertical component is roughly an order of magnitude stronger than the radial one.

To model the effects of passing stars on binary members, we employ the impulse approximation rather than directly integrating their passages[37]. This approximation is valid as long as the orbital period of the binary is much longer than the encounter timescale of the passing star. This holds for the vast majority of our stellar encounters. For instance, the shortest binary periods in our simulations are ∼20,000 years, whereas the encounter timescale of a passing star at 10,000 AU moving at 40 km/s is ∼1,000 years. (Such an encounter would hardly effect our tightest binary orbits, and the encounter timescale will be even faster for smaller impact parameters.) To set the encounter rates of our various stellar types (M, O, white dwarfs, etc.) we use a local observed mass function[38] combined with local velocity data for each stellar category[39]. Our stellar model assumes the local density of main sequence stars is 0.034 $M_\odot/pc^3$, and a mean dispersion of ∼42 km/s.

While MERCURY's wide binary integrator is excellent for modeling perturbations from a distant star as well as close encounters between planets, it cannot handle close encounters between a binary companion and a planet. This is not an issue for this set of simulations, however, for when the star comes close enough to our planetary system for star-planet encounters to arise, the planets will be destabilized already.

## 2.2 Generic Planetary System Simulations

In our second set of simulations, we model the evolution of artificially generated planetary systems orbited by a wide binary companion. To construct these planetary systems, we first choose a minimum planetary mass for each system that is between 0.5 and 5 $M_{\rm Jup}$. To choose this mass, we randomly draw from a $M^{-1.1}$ distribution[40]. Once a minimum mass is chosen, we next assign masses to the other two planets. To do this, we use the distribution of the observed mass ratios in known multi-planet systems where each planet is more massive than 0.5 $M_{\rm Jup}$. However, we clip this mass ratio distribution at 8, since larger ratios could potentially yield stellar-mass planets. Both the initial and final distributions of planetary masses are shown in Figure S1a. (Final distributions are used from the systems without a wide binary.) In addition, the distribution of mass ratios is shown in Figure S1b. As can be seen, the distributions do not evolve greatly. There is a small bias toward higher masses and mass ratios after systems have gone unstable. This is to be expected from planet-planet scattering[41].

Next, we assign orbits to each planet. First, a minimum semimajor axis is chosen randomly between 2 and 5 AU. The next planet's semimajor axis is set to a random value between 3.5–4.0 mutual Hill radii beyond the interior planet. The semimajor axis of the outermost planet is then set beyond the middle planet in an analogous manner. The mass order of the planets is randomized so that no correlation between mass and semimajor axis exists. With planetary mass and semimajor axis chosen, we now assign the rest of the orbital elements. Eccentricities are randomly drawn from a uniform distribution between 0 and 0.01, while inclinations are randomly drawn from a uniform distribution between 0 and 0.01 radians. Finally all other orbital elements are randomly drawn from a uniform distribution between 0 and $2\pi$. 200 planetary systems were constructed in this manner.

Finally, we assign wide binary companions to our planetary systems. We set all binary companion masses to 0.4 $M_\odot$, comparable to the typical mass of a star. To generate our binary orbital semimajor axes, we randomly select them from a distribution uniformly distributed in log-space between 1,000 and 30,000 AU[42, 43]. All other orbital elements are drawn from an isotropic distribution. 200 different binary orbits are generated in this manner.

To integrate these systems we use a modified version

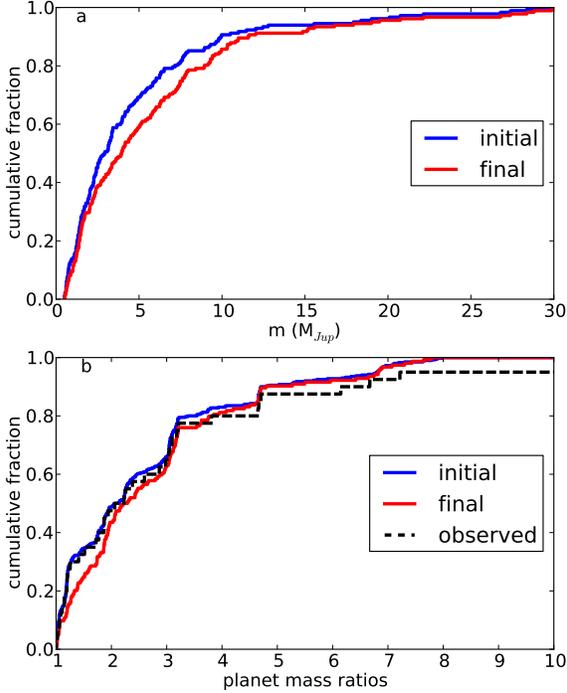

Figure S1. **a.** Cumulative distribution of planetary masses used in our planetary systems. Blue shows the initial distribution, and red shows the final distribution surviving after 10 Gyrs of evolution. **b.** Cumulative distribution of planetary mass ratios used in our planetary systems. Blue shows the initial distribution, and red shows the final distribution surviving after 10 Gyrs of evolution. The black is the observed distribution of mass ratios in systems containing multiple planets with $m \sin i > 0.5 \, \mathrm{M_{Jup}}$.

of MERCURY's wide binary integrator[31, 32]. Unlike the solar system simulations, we are very interested in the final architecture of these systems, so it is essential to accurately integrate these systems for the entire 10 Gyrs. There are two possible sources of inaccuracies. The first occurs when a planet is driven to very low pericenter, at which point integrations in democratic heliocentric coordinates are known to breakdown[44]. To guard against this, we set the initial timestep of our simulations to 1/60 of the innermost planet's orbital period (rather than the standard 1/20). Even with this small timestep, extreme eccentricities can cause rapid increases in numerical error. Therefore, if the innermost planet's eccentricity exceeds 0.7, we stop our symplectic integration and switch to a Bulirsch-Stoer integrator. This integrator is slower, but can accurately integrate very eccentric orbits. As mentioned in the previous section another source of potential error occurs during close encounters between planets and the binary companion. To guard against this, we also switch to a Bulirsch-Stoer integrator if the binary's pericenter comes within 100 Hill radii of the outermost planet's semimajor axis (typically of order 100 AU for our extended systems). Although there may be better integrator-switching criteria, this routine consistently triggers a Bulirsch-Stoer integration well before the symplectic integrator breaks down due to star-planet approaches or eccentric orbits.

It is well-known, however, that switching between integration schemes many times can also degrade the quality of an integration[45]. To prevent rapid switches between integration methods, once the Bulirsh-Stoer integrator is activated we wait 5 Myrs to determine if there are favorable conditions to resume a symplectic integration (low planetary eccentricities and large binary pericenter). If this is the case, we once again resume the sympectic integration using a timestep that is 1/60 of the innermost planet's current orbital period. Finally, we also cap the number of times that our algorithm can switch integration routines at 50. Depending on the initial conditions (isolated vs. binary star), this limit of 50 integration switches is only reached in 1–6% of runs.

To measure the effect of a binary star, each planetary simulation is run twice: once in isolation, and once surrounded by its binary companion. Like the solar system simulations, each run is integrated for 10 Gyrs.

With this modified integration routine, the numerical angular momentum error ($\frac{|dL|}{L_0}$) was held below $5 \times 10^{-6}$ for all simulations. Energy conservation was poorer. $\frac{|dE|}{E_0} < 5 \times 10^{-4}$ has been used as a benchmark in previous scattering simulations[46]. It was found that ~10% of our simulations did not initially meet this criterion. Ultimately, we found close encounters between planets to be the main source of energy error. A Bulirsch-Stoer integration is employed during these encounters, and the error tolerance of these integrations was initially set to 1 part in $10^{11}$. When we reran our simulations with a Bulirsch-Stoer tolerance of 1 part in $10^{15}$ only ~2% of our simulations had energy conservation poorer than $5 \times 10^{-4}$, with the worst cases being near $\frac{|dE|}{E_0} \simeq 10^{-2}$. While this is still not ideal, such a small fraction is unlikely to alter the global eccentricity distributions of the

17

main paper in any significant manner.

# 3 Supplementary Discussion

## 3.1 Initial Conditions of Generic Systems

As mentioned in the main text, the prescription we use for generating our generic planetary systems is likely overly simplistic. In particular, we have implied that all planetary systems form with exactly 3 Jovian-mass planets between 2 and ∼10 AU in a very unstable configuration. However, the point of these simulations is not to reproduce the exact dynamical evolution of planetary systems. Instead, they serve as a proof-of-concept demonstrating how the presence of a wide binary can alter planetary eccentricities. Although the results presented in the main paper are likely not unique to just the small corner of parameter space our initial conditions explore, we nevertheless believe it provides compelling support of our mechanism's importance. We have a dynamical model that reproduces the observed exoplanet eccentricity distribution around isolated stars. Depending on how radially extended our planetary systems are, we show the presence of a wide binary companion can alter this same dynamical model to yield an acceptable match to the observed exoplanets of known wide binary systems.

One other major assumption in our simulations is that we implicitly assume that planetary system formation is unaffected by the presence of a distant binary, since we integrate the same systems with and without a binary companion. While it is not certain if this assumption is correct, recent observations of protoplanetary disk dust emission around binary and isolated stars indicate that the disk emission of wide binaries and isolated stars are very similar[47]. Furthermore, the median pericenter of our initial distant binary orbits is ∼1600 AU, well beyond the observed edges of typical circumstellar disks, which have typical radii of ∼100 AU[48]. Indeed it has been demonstrated that much closer binaries are unlikely to disrupt protoplanetary disks[49].

### 3.1.1 Alternative Initial Conditions

It is important to mention that there are other known ways to excite the eccentricities of planets, which are affected by our choice of initial conditions. First, if our planetary systems are unstable and contain more than 3 Jovian planets we expect an even greater number of planet-planet scattering events to occur. Consequently, the surviving planets will have undergone many more close encounters with planets that are ultimately ejected, and we expect their final orbits to be more eccentric on average[50]. To verify this, we have also integrated 200 5-planet systems in isolation using the same mass distribution and spacing used in our 3-planet simulations. The eccentricity distribution of the innermost surviving planet in these simulations is shown in Figure S2a. We see that it is markedly more excited than the observed eccentricity distribution of planets around isolated stars. In fact, it's median eccentricity (0.44) is even higher than the median eccentricity of planets within wide binaries (0.39). Thus, inclusion of additional planets in systems within wide binaries could be an alternative explanation for the observed excitation of these systems.

In addition, another obvious way to alter the eccentricity excitation of planetary systems is by varying the inherent level of internal dynamical stability. In our 3-planet simulations presented in the main paper, we purposely place planets close to one another to trigger planet-planet scattering episodes almost immediately. This occurs in nearly every system (and our main paper's simulation analysis ignores the few systems that do not go unstable). If our planets were more widely-spaced we would expect a smaller fraction of systems to undergo planet-planet scattering episodes. Thus, the final eccentricity distribution of all systems would be markedly cooler. To demonstrate this, we have integrated an additional set of 200 3-planet systems in isolation. These systems have the same mass distribution as the 3-planet simulations used in the main paper. However, in this new set of simulations the planets are separated by 4.0–4.5 mutual Hill radii, as opposed to the 3.5–4.0 mutual Hill radii separations used in the main paper's simulations. The final eccentricity distribution of these more widely space systems is shown in Figure S2b. This is compared with the more packed isolated systems used in the main paper (set B1). We see that using more widely space planets drops the median eccentricity from 0.26 to 0.16. This is because only 2/3 of our more widely spaced systems undergo planet-planet scattering episodes, as opposed to 91% of the isolated 3-planet systems used in the main paper. Therefore, one more way to explain the heightened eccentricities observed among planets of wide binaries is that isolated planetary systems form in inherently more stable configurations.

The problem with both of the above explanations for

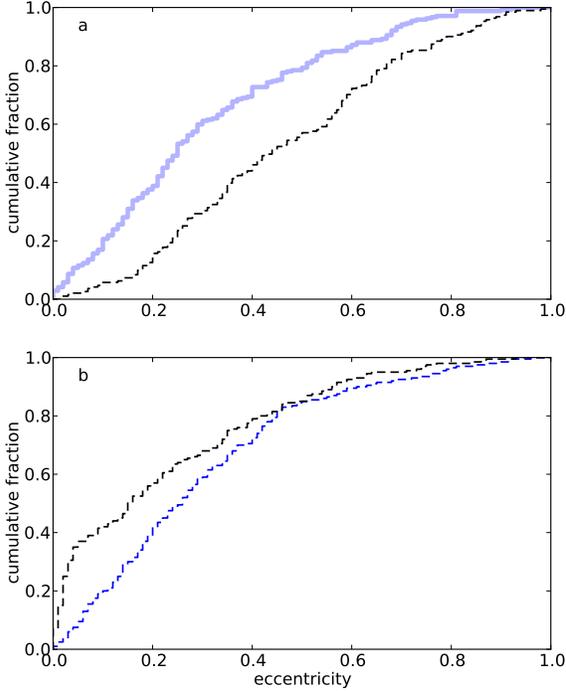

Figure S2. **a.** Cumulative eccentricity distribution for the innermost planets after 10 Gyrs of evolution in simulations beginning with five planets on unstable orbits (*dashed line*). Cumulative distribution of observed exoplanet eccentricities around isolated stars (*solid line*). Only planets with $a > 0.1$ AU and $m \sin i > 1$ M$_{\rm Jup}$ are included in the observed distribution. **b.** Final cumulative eccentricity distributions for the innermost planets in two different sets of 3-planet simulations. The planets are initially spaced by 4.0–4.5 mutual Hill radii in the first set (*black*) and by 3.5–4.0 mutual Hill radii in the second set (*blue*).

the observed excited eccentricities of planets in wide binaries is that they imply the process of planet formation is significantly different within very wide binaries compared to isolated stars. However, we show in our work that the members of very wide binaries spend most of their lives very far from one another. Phases where a binary orbit becomes very eccentric tend to be quite short-lived. Thus, we find it unlikely that a distant binary companion has a strong enough impact on planet formation to make the initial architecture of these planetary systems systematically different from isolated systems. Instead, we think that delayed instabilities caused by the time-varying nature of very wide binary orbits is a much more plausible explanation.

## 3.2 Planetary Semimajor Axes

We show the initial and final semimajor axis distributions in Figure S3. Here we see that the bulk of our planets are initially located between 2 and ∼10 AU. After internal instabilities occur, this distribution spreads out to both smaller and larger semimajor axes. The surviving inner planets of our simulations tend to perform most of the work ejecting and scattering other planets, and conservation of energy requires them to move to smaller semimajor axes. In addition, other planets can be scattered to large semimajor axes without actually being ejected, which populates the high-$a$ tail of our distribution.

It should be noted that the large majority of our simulated planets still orbit beyond the orbital distance of most known Jovian-mass planets (the observed median semimajor axis is ∼1.55 AU). We have intentionally done this to avoid prohibitive computing time for these simulations. However, when constructing the eccentricity distributions shown in Figure 3 of the main paper, we use only the innermost planet of each systems, as the outer planets often have decades-long orbital periods. If more distant planets reside in known exoplanet systems, it is likely that most have not yet been detected by radial velocity searches due to their longer orbital periods. In

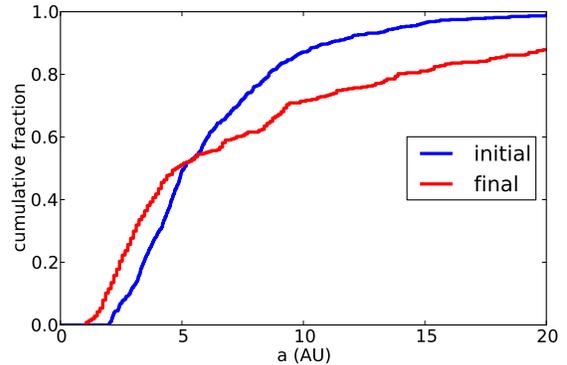

Figure S3. Cumulative distribution of planetary semimajor axes used in our planetary systems. Blue shows the initial distribution, and red shows the final distribution surviving after 10 Gyrs of evolution in isolation.

fact, even our innermost simulated planets tend to have more distant orbits than observed planets. However, it is still valid to compare the two groups' eccentricities, as there is currently little evidence that eccentricities generated by planet-planet scattering vary much with orbital distance[41].

## 3.3 Integration Times

One possible criticism of the work in the main paper is our choice to integrate our generic systems for 10 Gyrs. As stated above, our set of generic planetary systems is designed to be a "proof-of-concept," and the dynamical footprints of wide binary companions are strongest when our planetary systems are oldest. Furthermore, although determining stellar ages is fraught with uncertainty, it appears that Milky Way's thin disk began forming stars $\sim$10 Gyrs ago[33], and that most stars in the solar neighborhood are older than the Sun[34].

Nevertheless, some exoplanet-hosting stars are certainly younger than $\sim$10 Gyrs old, so it is important to show how our distribution of simulated planetary eccentricities evolves over time. In Figure S4, we show how the eccentricity distribution of our planetary systems evolves with age. In panel S4a, we plot the median eccentricity of the innermost planets of our simulated systems vs time. As can be seen, little eccentricity evolution takes place between 5 and 10 Gyrs. During the final 5 Gyrs, the median eccentricity of the innermost planets of our simulated systems only rises from 0.382 to 0.417. (Note that the median eccentricity of observed planets in wide binaries is 0.39.) In comparison, this value increases from 0.253 to 0.382 between $t = 10$ Myrs and $t = 5$ Gyrs. In addition, the bottom panel of Figure S4 demonstrates that a K-S test comparing our simulated planets with those in wide binaries returns only marginally different results during the final 5 Gyrs. Thus, using systems that are only 5 Gyrs old does not drastically weaken the agreement between our simulated planetary eccentricities and observed exoplanet systems. Furthermore, one could easily imagine accelerating the eccentricity evolution seen in Figure S4 by using more radially extended planetary systems. In addition, eccentricity evolution could be sped up in a denser galactic environment where perturbations are stronger (see Section 3.10 for a discussion).

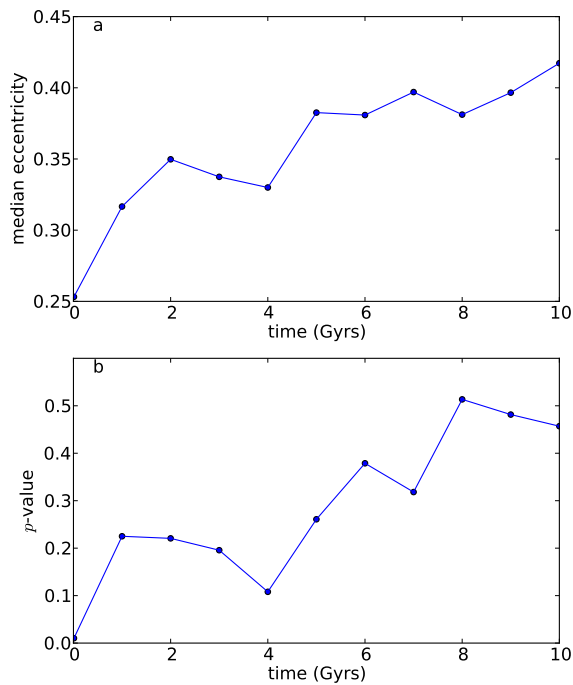

Figure S4. **a.** Plot showing the evolution of the median eccentricity of the innermost planets in our simulated 3-planet systems as a function of time (leftmost data point corresponds to $t = 10$ Myrs). **b.** Time evolution of the p-value of a K-S test when comparing our simulated planetary eccentricities with those of observed planets in very wide binaries.

## 3.4 Further Notes on Observed Eccentricities

The catalog of extrasolar planets is far from complete, and there exist well-known biases in the observed data (detection of short-period, massive planets). It is important then to consider if these biases might explain the heightened planetary eccentricities seen within wide binary star systems, especially since we use this trend to support and constrain our numerical work. Because many early radial velocity exoplanet searches looked to our own solar system as the archetype planetary system, these campaigns originally targeted isolated stars and tended to avoid binary systems.

Most exoplanet-hosting binary stars are located relatively close to the Sun (this enables the detection of

the stellar companions). Consequently, most of these exoplanets are discovered with the radial velocity technique, since these campaigns tend to target nearby stars. Therefore, when comparing planets around isolated stars to planets within wide binaries, we only consider planets discovered via radial velocity, as listed on exoplanets.org. This also limits our biases to just those associated with this detection technique.

Works cataloging exoplanets within binary star systems can be broadly divided into two groups: 1) studies that perform follow-up observations on known planet-hosting stars to specifically search for unknown stellar companions[51, 52, 53, 54, 55], and 2) studies that attempt to cross-match the catalog of known exoplanet host stars with catalogs of known stellar multiples or common proper motion pairs[55, 56]. Thus, most exoplanet-hosting binaries were originally mistaken for exoplanet-hosting isolated stars. Only after exoplanets were discovered around these stars were they realized to be members of multi-star systems. Although the catalog of exoplanet-hosting binaries is subject to observational biases and is compiled from a multitude of different works, this catalog is ultimately a subset of detections from the same campaigns that have yielded the more extensive catalog of exoplanets around isolated stars with the same detection techniques. Therefore, it is not obvious how the excited eccentricities of exoplanets within wide binaries could arise from an observational bias not present in the catalog of exoplanets that orbit isolated stars.

### 3.4.1 Eccentricity Discrepancy as a Function of Planetary Mass

In the main text, we only consider observed exoplanets with masses above 1 $M_{\rm Jup}$. This is done because recent analyses of exoplanet eccentricities show a well-defined division between the eccentricities of planets more massive than ∼1 $M_{\rm Jup}$ and those less massive than 1 $M_{\rm Jup}$[57, 41]. While earlier works placed the dividing mass at even higher exoplanet values of ∼4 $M_{\rm Jup}$[58, 59], our limited catalog of exoplanets within wide binaries would quickly be depleted if we increased our mass cutoff. However, it is worth considering how the eccentricity distribution of exoplanets within wide binaries compares to isolated exoplanet systems when we instead decrease the minimum planetary mass that we consider. In the main text, we find a K-S test returns a $p$-value of 0.006 when we use a minimum planetary mass of 1 $M_{\rm Jup}$. When we instead consider all planets more massive than 0.25 $M_{\rm Jup}$ the $p$-value increases to 0.04, and

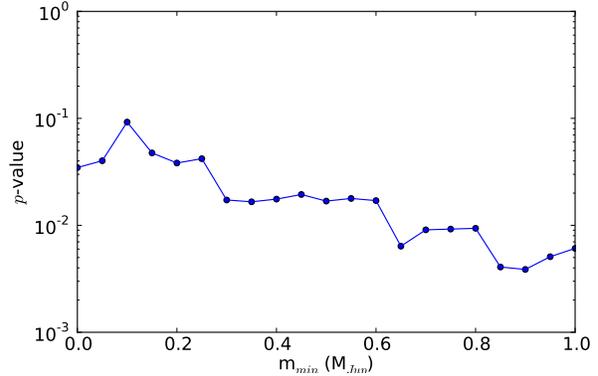

Figure S5. We use a K-S test to compare observed planetary eccentricities around isolated stars with planetary eccentricities found within wide binaries. For this test, we only consider planets with masses greater than a minimum mass, $m_{min}$. The p-value of a K-S test comparing these two eccentricity distributions is plotted as a function of the minimum planetary mass we consider.

the differences between the two exoplanet distributions are less severe.

A plot of how the $p$-value of a K-S test varies as a function of our minimum planetary mass is shown in Figure S5. As can be seen, this value increases with decreasing planetary mass, climbing above 0.01 when we consider planets more massive than 0.6 $M_{\rm Jup}$ and finally settling at 0.035 when all planets are considered. Although the statistics are poor (we only have 12 exoplanets whose masses are between 0 and 1 $M_{\rm Jup}$ residing within wide binaries), this suggests that the discrepancy between the two eccentricity distributions is confined to high-mass planets. This potentially supports our explanation that binary-triggered instabilities cause this eccentricity excitation. If a planetary system within a wide binary happens to avoid a binary-triggered instability, then the evolution of planetary orbits should be identical to those around isolated stars. On the other hand, if the wide binary evolves to a low pericenter and greatly excites the planetary orbits, we expect only high-mass planets will preferentially survive, while the low-mass ones will be preferentially ejected. Thus, the systems most harassed by binaries will be largely devoid of low-mass planets, while those unaffected by their binary companion may host planetary systems similar to isolated systems.

### 3.4.2 Catalog Contamination by Undiscovered Binary Systems

As direct imaging studies have demonstrated[53, 54], planet-hosting binaries are typically mistaken for isolated stars before a stellar companion is detected. This then raises the prospect that planet-hosting binaries pollute the catalog of isolated planet-hosting stars and distort their planetary orbital distributions. However, there is a limit to the level of contamination. As mentioned above, many of the original exoplanet searches avoided known binaries and targeted stars thought to be isolated. Consequently, our catalog of isolated planet-hosting stars should have a lower frequency of wide binaries than a random sampling of field stars, and we can use the observed rate of wide binaries in the field as a maximum estimate of possible contamination. Studies using common proper motion pairs [60] and angular two-point correlation functions[61] have found that binaries with $a_* \gtrsim 10^3$ AU account for $\sim$10% of all stars in the solar neighborhood. We can then estimate the maximum effect of binary contamination by assuming that 10% of all isolated planet-hosting are actually binaries. To estimate this, we randomly sample the eccentricity distribution of exoplanets within wide binaries to generate a list of "fake" planetary eccentricities that is 10% as extensive as our catalog of planets around isolated stars. Adding this list of fake eccentricities to our real catalog only changes the median eccentricity of planets around isolated stars from 0.25 to $\sim$0.26. Thus, the effect of binary contamination seems to be negligible.

However, there is also the possibility that isolated stars once had distant stellar companions that were eventually stripped off by impulses from passing stars. These previously binary stars would then be included as isolated stars. To correct for this effect in our simulated eccentricity distributions of planets within wide binaries, we make sure to only analyze simulated planets orbiting stars that still retain a binary companion. We cannot correct for contamination among our observed isolated systems, though. Once again, we can estimate the maximum effect of this contamination. In our simulations of wide binaries, we see that only for $a_* \gtrsim 10^4$ AU are the majority of binaries ionized over the age of the Milky Way's thin disk (see discussion of Figure S8 and Section 3.9). Further, upon examining Figure 2 of the main paper, we see that the rate of binary-triggered instabilities begins to diminish beyond $a_* \sim 20,000$ AU because these binaries are ionized very rapidly. Thus, the wide binaries most likely to affect planets and then get ionized later have initial semimajor axes between 10,000 and 20,000 AU. If $\sim$10% of stars form with wide binary companions, and if wide binary semimajor axes follow an Opïk distribution, which is distributed uniformly in log-space, then this range of binary semimajor axes would correspond to 2-3% of all stars. Therefore, by the result of the previous contamination analysis, this effect should be negligible as well.

### 3.5 Evolution of Isolated Systems

While the main text focuses on the evolution of our simulations that include a binary companion, it is important to also document the evolution seen in our isolated 3-planet systems. Most of these systems undergo internal instabilities very quickly. (Note that we consider an instability to occur when a planet is lost from the system.) In Figure S6, we plot the fraction of systems that have gone unstable as a function of time. One sees that the large majority (81%) of our systems become unstable before $t = 10$ Myrs. It should also be noted that 9% of our isolated planetary systems are stable for the entire 10-Gyr integration. We assume that planet-planet scattering is an ubiquitous process in planetary systems, and because of this, we do not include these systems in Figure S6 or in the eccentricity distributions (both isolated and wide binary) in Figure 3 of the main paper.

In the final part of the main text, we use the planetary systems that have collapsed to 2-planet systems to argue that wide binaries affect planetary systems extended beyond 10 AU much more strongly than those

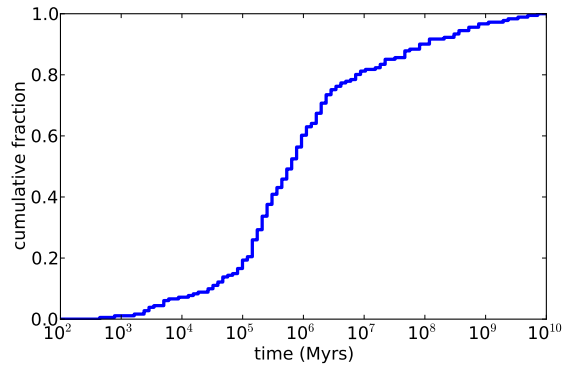

Figure S6. Cumulative distribution of the times that isolated 3-planet systems undergo their first dynamical instability. Systems that remain stable for 10 Gyrs are excluded from this distribution.

confined inside 10 AU. To support this claim, we examine the instability rate of similar systems in our isolated batch of simulations. At $t = 10$ Myrs, we examine all of the 2-planet systems in our isolated runs. For those with an outer planet beyond 10 AU, 16% of our systems experience another instability at a later time. This is a very similar rate to our systems confined inside 10 AU (19%). For comparison, the instability rate is much higher for our extended 2-planet systems that include a wide binary. 60% of these systems experience additional instabilities. Meanwhile our compact 2-planet systems with wide binaries have instability rates closer to the isolated cases, with only 25% of these systems experiencing instabilities after $t = 10$ Myrs. Thus, it seems that wide binary companions preferentially destabilize extended planetary systems.

Finally, it should be noted that these instability statistics only apply to the simulations used to construct Figure 3 in the main text (simulation sets B1 through B3). In Figures 1 and 2 we integrated the four giant planets of the solar system in the presence of a very wide binary (simulation set A). Obviously, this planetary configuration is stable indefinitely in the absence of binary perturbations.

## 3.6 Instability Rate as a Function of Planetary Semimajor Axis

In Figure 3 of the main text, we show that planetary systems extended beyond 10 AU are much more likely to suffer disruption from their binary companion than systems with planets all confined within 10 AU. To better quantify this trend, we perform an additional set of simple simulations with more well-tuned initial conditions. These simulations consist of 1,000 runs with 10 different systems of two 1 $M_{\rm Jup}$ planets (100 runs per planetary system). In each planetary configuration, the planets are started on nearly circular, coplanar orbits separated by 6 mutual Hill radii (a stable orbital configuration). The parameter varied between configurations is the semimajor axis of the outer planet, with values ranging between 2.5 and 100 AU. These systems are then integrated in the presence of a 0.4 $M_\odot$ companion whose orbital semimajor axis is drawn from a uniform distribution in log-space between 1,000 and 30,000 AU. All other elements are drawn from isotropic distributions.

Our systems are then integrated for 10 Gyrs or until a planet is lost from the system due to a binary-triggered instability. The fraction of planetary systems

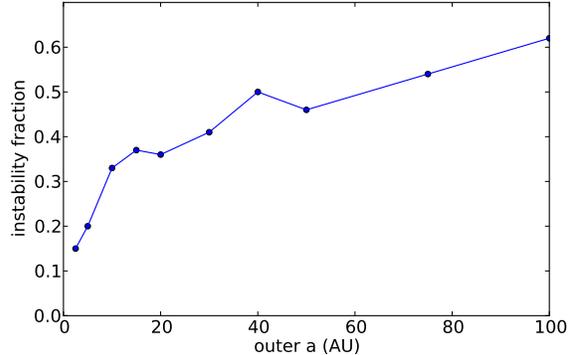

Figure S7. Plot of the fraction of 2-planet systems that lose one or more planets due to binary-triggered instabilities as a function of the initial semimajor axis of the outermost planet.

that have gone unstable is plotted as a function of the outer planet's semimajor axis in Figure S7. As can be seen, the instability fraction for planetary systems extending to 2.5 AU is ∼15% but more than doubles by 10 AU and exceeds 50% by 40 AU. Although this essentially just measures the rate that the outermost planet is excited by a binary, this excitation often cascades inward in packed planetary systems, exciting the orbits of observable planets [62]. This process is illustrated in Figure 1 of the main paper, where Uranus is actually the first planet ejected via binary perturbations even though Neptune's orbit is most strongly perturbed. Again, the observed eccentricity excitation in known planets of wide binaries combined with the rapid rise in instability rates between 2.5 AU and ∼10 AU shown in Figure S7 suggests that most systems of giant planets extend to tens of AU.

## 3.7 Lost Planet Fates

Because we begin our generic planetary systems in unstable configurations, many planets are lost throughout the course of our 10-Gyr integrations. In our isolated batches, 12% of our lost planets collide with the central star ($r < 0.01$ AU), 15% suffer collisions with other planets, and the remaining 73% are ejected from the system ($r > 10^5$ AU). In our wide binary batch of simulations, the fractional breakdown is similar, with 14% of lost planets colliding with the star, 13% colliding with other planets, and 73% ejected. It should also be noted that many more planets are lost in our wide binary

23

simulations (60%) vs. our isolated simulations (41%). The enhanced planetary loss rate in our simulations with wide binaries is why 70% more of these systems experience a planetary collision with the central star when a wide binary is included, even though the chance of collision is still relatively small. This enhanced loss rate is not surprising, as previous works have shown that close stellar flybys can greatly increase planetary ejections[63, 64, 50]. In tighter binaries, it has been shown that a small fraction of these ejected planets may be exchanged with the other binary member[65]. Unfortunately, our simulations may not be capable of measuring this statistic, since MERCURY's wide binary symplectic algorithm always assumes the planets are in orbit about the primary. We intend to study the phenomenon of planetary exchange further in the future.

### 3.8 Planet Stripping

As mentioned in the main text, $\sim$20% of our simulated systems are completely stripped of planets by their wide binary companion. In these cases, the binary orbit reaches such a low pericenter that even the innermost planet is destabilized. In most of our stripping cases, the final surviving planet is driven into the star via Kozai oscillations. This occurs for 64% of our stripped systems. In the remaining cases the last planet is dynamically unbound from its host star via strong perturbations from the close-passing binary companion. Based on these results we predict that up to 1 in 5 wide binary stars possess no planetary systems at all, even though they may have formed planets early in their history. However, as discussed in the main text, for the planets driven into their stars by Kozai resonances, tidal dissipation during pre-collision close passages between the host star and planet may allow some of these planets to survive[66, 67].

### 3.9 Binary Ionization

As mentioned in the main text, the impact of our most distant binaries on planetary evolution is limited by the fact that they are quickly ionized by perturbations from the local galactic environment. In fact, of our 200 3-planet simulations, $\sim$1/4 of our systems lose their binary companion after 10 Gyrs of evolution. Because these planetary systems would be classified as isolated systems after binary ionization, planetary eccentricities from these ionized systems are not included in Figure 3 of the main text.

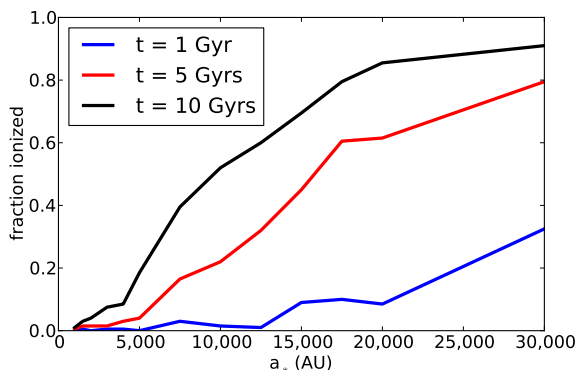

Figure S8. Fraction of binaries that have been ionized at $t = 1$ Gyr (blue), 5 Gyrs (red), and 10 Gyrs (black). This fraction is plotted as a function of the initial binary semimajor axis.

Because our solar system simulations integrated large numbers of binary stars at specific initial semimajor axes, we can use these simulations to study the ionization rate of wide binaries in the local galactic environment. In Figure S8, we plot the fraction of binaries that become ionized as a function of their initial semimajor axis for various times. As can be seen, our population of $a_* = 30,000$ AU binaries is quickly eroded. After 1 Gyr of evolution $\sim$1/3 of these binaries have been lost, and this fraction exceeds 90% after 10 Gyrs. $a_* \simeq 10,000$ AU is roughly the point at which binary ionization plays a large role in the evolution of these systems. Over half of systems with initial semimajor axes beyond this value are ionized in 10 Gyrs. Inside of this, the ionization rate continues to fall, with $< 10\%$ of binaries ionized inside $a_* = 4,000$ AU.

### 3.10 A Variable Galactic Environment

As mentioned above, we choose to hold the both the population of passing stars and the strength of the Galactic tide fixed for our simulations. This is done to keep our simulations simple and limit the number of parameters we vary. However, it has been shown recently that most stars within spiral galaxies radially migrate over large galactocentric distances on Gyr timescales[68, 69]. In addition, an outward migration appears likely for stars with Sun-like kinematics[35]. This suggests that many wide binaries in the present day solar neighborhood once inhabited denser regions of the Galaxy where external perturbations from passing stars and the Galaxy's

tide would have been stronger. This would decrease the timescale on which wide binary orbits evolve, and consequently, it is likely that wide binary companions would have an even greater impact on the planetary system evolution than our simulations suggest in the main text.

As a simple demonstration of this effect, we rerun our solar system simulations (set A in the main paper) that have a 0.1 $M_\odot$ wide binary companion. In the original simulations, 35.7% of our planetary systems ejected at least one planet. When we double the local galactic disk density and stellar density we find that the percentage of destabilized planetary systems increases to 44.5%. Finally, if we double the local density again this percentage climbs to 47.4%. In a study of the migratory history of Sun-like stars, it was found that the mean density of the time-integrated local galactic environment is typically more than twice as high as the present day solar neighborhood[35], supporting the idea that the simulations in the main text represent a conservative estimate of the importance of wide binary stars on planetary systems.

## 3.11 Tighter Binaries

For our simulations presented in the main paper, we do not consider binaries with semimajor axes below $10^3$ AU. However, in Figure S9 we extend Figure 2 of the main paper down to binary semimajor axes of 500 AU. Figure S9a indicates that the fraction of planetary systems destabilized by binaries becomes extremely high ($> 90\%$) in binaries with semimajor axes below $10^3$ AU. Based on these results, it is even more surprising that only planets of very wide ($a_* > 10^3$ AU) are observed to have excited eccentricities. Such a trend seems to contradict this figure. However, Figure S9 may be misleading for several reasons. First, in each of the 2800 simulations that comprise this figure, we use the solar system's four giant planets to measure the rate that binaries trigger planetary instabilities. This assumes that the presence of a binary does not affect planet formation enough to yield systematically different types of initial planetary systems around binaries of different separations. Further, by using planetary systems like the solar system, we are assuming that binaries form initial planetary systems like those around isolated stars like our Sun. For very wide binaries, this is a reasonable assumption, since we show in the main paper that the periods in which two binary member stars closely approach each other are rare and brief, and therefore unlikely to alter the relatively short epoch of planet for-

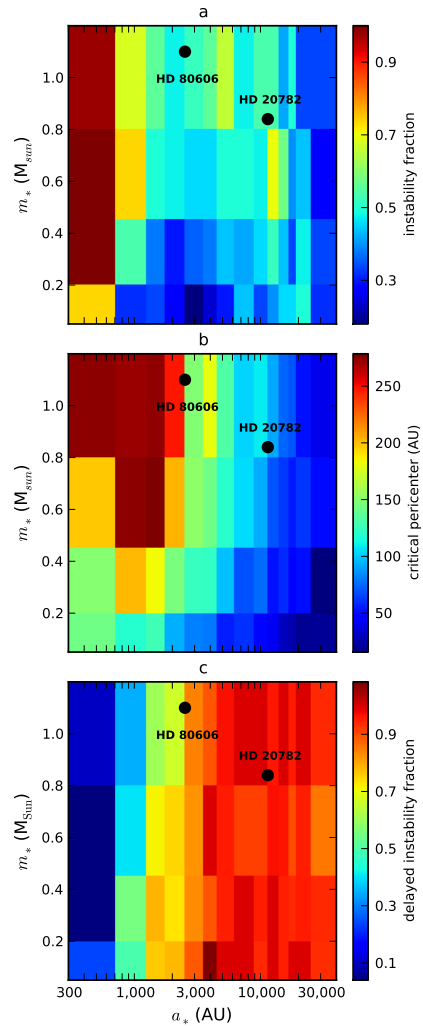

Figure S9. Alternate version of Figure 2 of the main paper extended down to binary semimajor axes of 500 AU. **a.** Map of the fraction of systems that lost at least one planet via instability. Binary mass is plotted on the $y$-axis, while the $x$-axis marks binary semimajor axis. **b.** The median binary pericenter below which an instability is induced in the planetary system as a function of binary mass and semimajor axis. **c.** The fraction of all binary-triggered instabilities that occurred after 100 Myrs as a function of binary mass and binary semimajor axis. In each panel black data points mark the masses and presumptive semimajor axes of the HD 80606 and HD 20782 binaries, which host the two most eccentric known planetary orbits[59, 80].

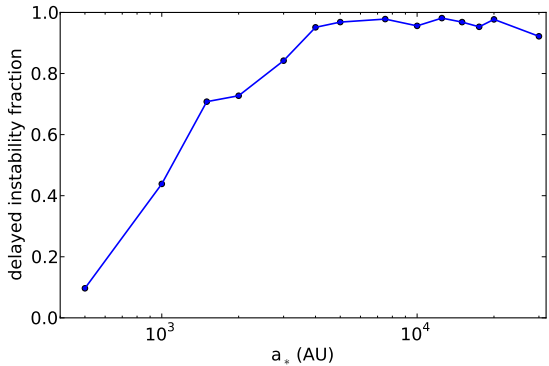

Figure S10. Fraction of binary-triggered instabilities from Figure S9 that are delayed until after $t = 100$ Myrs as a function of binary semimajor axis.

mation around each star. However, for binaries with $a_* \lesssim 10^3$ AU this becomes a much more questionable assumption. For our binaries with $a_* = 500$ AU, the median initial pericenter of their orbits is only ~150 AU. Thus, they potentially have a much more significant impact on planet formation, and it's not clear that we should even expect solar system-like architectures to arise in binaries with $a_* \lesssim 10^3$ AU. Perhaps then it is not very surprising that such planetary systems are almost always immediately destabilized when placed in binaries with $a_* < 10^3$ AU.

The constant significant gravitational interaction between members of tight binaries is evidenced in Figure S9c. Here we see that the timescale for binaries to trigger planetary system disruptions becomes much shorter below $a_*$ of 1,000–2,000 AU. In Figure S10, we sum these simulations over all binary masses to see what fraction of instabilities occur well after planet formation ($t > 10^8$ yrs) solely as a function of binary semimajor axis. Here we see that for binary semimajor axes beyond ~3,000 AU at least ~95% of all instabilities are delayed. Inside $a_* < 10^3$ AU, however, this fraction plummets to just under 10%, implying that most systems are immediately unstable. If these planetary systems are instantly unstable, we may actually expect planet formation to be altered enough to prevent formation of solar system-like planetary architectures in such binaries.

Another reason the low-$a_*$ region of Figure S9 may be inaccurate is because of the initial distribution of binary orbital elements we assume in our simulations. It is thought that binaries with $a_* \lesssim 10^3$ AU have a funda-

mentally different formation process than binaries with $a_* \gtrsim 10^3$ AU. For binaries with $a_* \gtrsim 10^3$ AU, it is largely thought the formation mechanism is dynamical capture either due to 3-body interactions or star cluster dispersal[70, 71, 72]. Such processes yield an isotropic distribution of binary eccentricities and other orbital elements[71], which is what we assume in our simulations' initial conditions. On the other hand, tighter binaries ($a_* \lesssim 10^3$ AU) are thought to form via fragmentation of collapsing cloud cores[73]. In this scenario, it is less clear what the resulting distribution of orbital elements should be. Indeed, studies of imaging polarimetry of young binaries indicate that the circumstellar disks of binary members are coplanar to within 20° for binary separations between 200–1,000 AU[74, 75]. This would imply that the binary orbital planes must also be nearly coplanar with the circumstellar disks, or else these disks would rapidly become misaligned.

Thus, if inclinations between tighter binary orbits and planetary disks are colder, the eccentricities may be as well. This could also explain why the observed planetary eccentricities with tighter binaries are colder than those found in binaries with $a_* > 10^3$ AU. Such an orbital distribution is much less likely to result in planetary system disruptions. To demonstrate this, we integrate nearly coplanar ($i < 20°$), nearly circular ($e < 0.1$) binary companions with masses of 0.4 $M_\odot$ about the four giant planets of the solar system for 10 Gyrs.

We perform 20 integrations at binary semimajor axes of 300, 500, 750, 1,000, 3,000, 5,000, and 10,000 AU

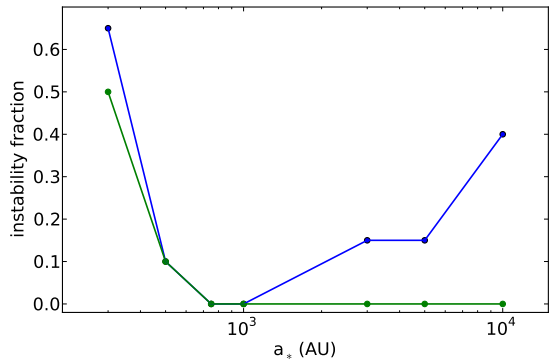

Figure S11. Fraction of systems that lose at least one planet due to a binary-triggered instability as a function of binary semimajor axis. Systems are run in the presence of galactic perturbations (*blue*) and with perturbations shut off (*green*).

for a total of 140 runs. The fraction of planetary systems that have undergone an instability resulting in at least one planetary ejection is plotted as a function of binary semimajor axis in Figure S11. In this plot we see that binaries with $a_* \sim 300$ AU are still quite disruptive to systems like our solar system. However, between $a_* \sim 500$ AU and $a_* \sim 3,000$ AU the planetary systems are relatively unaffected by the presence of the binary. Beyond $a_* \sim 3,000$ AU, the disruption rate picks up again, as perturbations from the Galactic tide and passing stars are powerful enough to completely transform the initial eccentricity distribution. (It should be noted that our initial cold eccentricity distribution largely suppresses binary-triggered disruptions for binaries with $1,000 < a_* < 5,000$ AU compared to the results of the main paper. This is because the timescale for the Galactic tide drive a circular orbit to an extreme eccentricity is longer than our integration time. Planetary system disruption in these binaries typically results instead from initially moderately eccentric binary orbits driven to extreme eccentricities.)

Although these simulations' initial conditions are idealized, they illustrate an important point. There are zones of binary orbital parameter space that allow certain classes of planetary systems to remain stable indefinitely in tighter binaries. When the binary separation is increased, however, galactic perturbations drive binaries far from their initial orbital configurations, regularly disrupting the same planetary systems. To further illustrate this, we rerun our simulations with galactic perturbations turned off. The planetary instability rates found in these simulations are also shown in Figure S11. As can be seen, without galactic perturbations, the instabilities occurring within the largest binary semimajor axes have disappeared.

We admit that our explanations offered for the unexcited planetary eccentricities observed in tighter binaries are not fully tested and are a bit speculative. However, there are already many works focused on planetary dynamics for binary separations at or below $\sim 10^3$ AU[76, 77, 66, 49], while our work is among the first to focus on planets within the widest binaries. Planetary dynamics within tighter binaries will be a topic for our future research, but a fully detailed study of this problem is beyond the scope of the present work.

## 3.12 Wide Binary Formation

In the main paper, we demonstrate that the orbital eccentricities of observed giant exoplanets with wide binary companions are statistically higher than those without stellar companions. We argue that this is due to the dynamical evolution of the wide binary systems under the influence of other passing stars and the Galactic tide. However, the exact process of wide binary formation remains poorly understood, and here we consider the possibility that the process of wide binary formation yields more excited planetary systems.

Several recent works have proposed that wide binaries form during the dissolution of star clusters[71, 70, 72]. In such scenarios, the two stars are already tenuously bound during cluster dissolution and are "frozen out," or they become bound to one another as they escape the dissolving cluster potential. These formation mechanisms seem unlikely to excite planetary eccentricities in these systems because the two member stars should only rarely closely encounter one another during the formation process. Such encounters are not required in these formation scenarios, and they would only arise if a system forms with an initial binary orbit that is extremely eccentric. Such initial orbits are unlikely for any plausible binary orbit eccentricity distribution, even isotropic.

However, another new formation mechanism has been proposed suggesting that most wide binaries are born as triple star systems[78]. In this mechanism, the triple system quickly goes unstable. Typically this would result in the ejection of one of the stars. However, if the instability occurs quickly enough the potential of the star-forming molecular cloud core can instead trap a star in a distant orbit before it is ejected. What results is a "binary" that is actually composed of one well-separated star and two other stars in a much tighter orbit about each other. Unlike the previous binary formation mechanisms, close encounters between member stars could be common as triple star systems go unstable in this scenario. Thus, it seems possible that any planetary systems that are present could be significantly perturbed during these initial triple star system instabilities.

Perhaps then, the excited planetary eccentricities we discuss in the main paper are actually a signature of wide binary formation from triple stars rather than a signature of the subsequent dynamical evolution of wide binaries. There are two reasons we do not believe this is the case, however. First, if most wide binaries form from triple systems this requires most known wide binaries to actually be comprised of three stars (an isolated star and a tight binary). In the main paper, we analyze the orbits of 20 known exoplanets with wide binary companions.

27

Of these 20, only 4 reside in systems where the distant binary companion is known to be a much tighter binary itself. While these are well-studied systems, it is possible that the tight binary has been mistaken for a single star in some of the more distant systems. However, systems within ∼30 pc should indeed have all stellar members detected[78]. Of the known planet-hosting wide binaries, six are actually within ∼30 pc, and only one of these systems (16 Cygni) is known to be actually be comprised of three stars, which suggests that most of the more distant planet-hosting wide binaries also only have two stellar components. Thus, the predominance of conventional "two-star" wide binaries in our planetary sample suggests that triple stars do not yield most wide binaries that host giant planets. (It should also be noted that these 4 planets are not particularly eccentric members of the overall sample. Only one has an eccentricity that exceeds the overall sample's median.)

A second reason we believe that instabilities of triple star systems do not explain the heightened eccentricities of exoplanets within wide binaries relates to the timing of triple star instabilities. Because the potential of a molecular cloud core is necessary to prevent the distant star from being ejected, these instabilities must occur very early in the lifetime of stars, and indeed most are predicted to occur in the protostellar stage[78]. Consequently, it is not at all clear that the stars involved in this process have formed planets yet when the instability occurs. If planetary systems do not form until after this instability then we expect the distant stellar companions to have little effect on the final orbital eccentricities of the planetary systems. Only later when galactic perturbations drive the stellar orbit to very low pericenter as described in the main paper can the distant companion excite or disrupt the planetary system.

## 4 Supplementary Table

In Figure 3 of the main paper, we construct two distribution functions of observed exoplanet eccentricities within binary star systems. Those planets are listed on the next page in Table S1 along with their orbits and masses as well as a description of their binary star system. (It should be noted that in a few cases, the distant binary companion is itself actually a much tighter binary system. For large separations, the potential of this binary will closely resemble a point mass.) Although only the projected separations of most binaries are known, the most probable semimajor axis can be calculated assuming an isotropic distribution of orbital elements[79]. This is just 1.26 times larger than the projected separation, and we use this value to divide our sample into planets of very wide binaries ($a_* > 10^3$ AU) and planets of tight binaries ($a_* < 10^3$ AU). Based on this division, we have 20 planets within very wide binaries and 23 planets that reside in tight binaries. As stated in the main text, our analysis does not include planets with $m \sin i < 1$ $M_{\rm Jup}$ or $a < 0.1$ AU. These cuts exclude ∼43% of all known planets within binaries.

| Name | Projected Stellar Separation (AU) | $a_*$ (AU) | $m \sin i$ ($M_{\rm Jup}$) | $a$ (AU) | $e$ |
|---|---|---|---|---|---|
| HD 38529c | 12,040 | 15,180 | 17.7 | 3.695 | 0.36 |
| HD 20782b | 9,080 | 11,550 | 1.8 | 1.36 | 0.92 |
| HD 40979Ab | 6,400 | 8,070 | 3.83 | 0.855 | 0.269 |
| HD 147513b | 5,360 | 6,760 | 1.21 | 1.32 | 0.26 |
| HD 222582b | 4,750 | 5,990 | 7.75 | 1.35 | 0.76 |
| HD 125612b | 4,750 | 5,990 | 3.0 | 1.37 | 0.46 |
| HD 125612d | 4,750 | 5,990 | 7.2 | 4.2 | 0.28 |
| HD 213240b | 3,900 | 4,920 | 4.72 | 1.92 | 0.421 |
| GJ 777b | 3,000 | 3,780 | 1.56 | 4.01 | 0.313 |
| HD 7449b | 2,960 | 3,730 | 1.1 | 2.3 | 0.82 |
| HD 7449c | 2,960 | 3,730 | 2.0 | 4.96 | 0.53 |
| HD 89744b | 2,460 | 3,100 | 8.58 | 0.934 | 0.677 |
| HD 219449b | 2,250 | 2,840 | 2.9 | 0.3 | 0.0 |
| HD 80606b | 2,000 | 2,520 | 3.9 | 0.453 | 0.9336 |
| 30 Arietis Bb | 1,500 | 1,890 | 9.88 | 0.995 | 0.289 |
| 55 Cancri d | 1,065 | 1,340 | 3.82 | 5.74 | 0.014 |
| 11 Comae b | 995 | 1,250 | 19.4 | 1.29 | 0.231 |
| 16 Cygni b | 860 | 1,080 | 1.68 | 1.681 | 0.681 |
| HD 142022b | 820 | 1,030 | 4.47 | 2.93 | 0.53 |
| GJ 676b | 800 | 1,010 | 4.9 | 1.82 | 0.326 |
| HD 178911b | 785 | 990 | 7.35 | 0.345 | 0.139 |
| Upsilon And c | 750 | 945 | 13.98 | 0.832 | 0.224 |
| Upsilon And d | 750 | 945 | 10.25 | 2.53 | 0.267 |
| Upsilon And e | 750 | 945 | 1.059 | 5.24 | 0.005 |
| HD 188015b | 684 | 862 | 1.5 | 1.203 | 0.137 |
| HD 196050b | 510 | 643 | 2.9 | 2.54 | 0.228 |
| HD 132563 Bb | 397 | 500 | 1.49 | 2.6 | 0.22 |
| HD 65216b | 253 | 319 | 1.21 | 1.37 | 0.41 |
| HD 156846b | 250 | 315 | 10.45 | 0.993 | 0.8472 |
| HD 27442b | 240 | 302 | 1.56 | 1.271 | 0.06 |
| HD 28254b | 238 | 300 | 1.16 | 2.15 | 0.81 |
| GJ 667 Cc | 204 | 257 | 4.54 | 0.123 | 0.27 |
| Gamma Leo b | 170 | 214 | 8.78 | 1.19 | 0.14 |
| HD 19994b | 151 | 190 | 1.68 | 1.42 | 0.3 |
| HD 195019b | 150 | 189 | 3.69 | 0.1388 | 0.0138 |
| HD 142b | 138 | 174 | 1.31 | 1.045 | 0.26 |
| HD 114762b | 130 | 164 | 11.68 | 0.363 | 0.3359 |
| HD 177830b | 97 | 122 | 1.49 | 1.22 | 0.009 |
| GJ 3021b | 68 | 86 | 3.37 | 0.49 | 0.511 |
| HD 196885b | 24 | 30 | 2.96 | 2.6 | 0.48 |
| Gl 86b | 21 | 26 | 3.91 | 0.113 | 0.0416 |
| HD 41004 Ab | 20 | 25 | 2.56 | 1.7 | 0.74 |
| Gamma Cep Ab | 20 | 25 | 1.6 | 2.044 | 0.115 |

Table S1. List of the planets residing in binaries which are used to construct the observed distribution functions in Figure 3 of the main paper. Columns from left to right are: planet/star name, binary projected separation distance, most likely binary semimajor axis, planetary mass, planetary semimajor axis, and planetary eccentricity. Systems are listed in order of decreasing binary separation.




\begin{thebibliography}{10}

\bibitem{roe12}
{Roell}, T., {Neuh{\"a}user}, R., {Seifahrt}, A., and {Mugrauer}, M.
\newblock {Extrasolar planets in stellar multiple systems}.
\newblock {\em Astron. Astrophys.}{ \bf 542}, A92 (2012).

\bibitem{hegras96}
{Heggie}, D.~C. and {Rasio}, F.~A.
\newblock {The Effect of Encounters on the Eccentricity of Binaries in
  Clusters}.
\newblock {\em Mon. Not. R. Astron. Soc.}{ \bf 282}, 1064--1084 (1996).

\bibitem{jiatre10}
{Jiang}, Y.-F. and {Tremaine}, S.
\newblock {The evolution of wide binary stars}.
\newblock {\em Mon. Not. R. Astron. Soc.}{ \bf 401}, 977--994 (2010).

\bibitem{oort50}
{Oort}, J.~H.
\newblock {The structure of the cloud of comets surrounding the Solar System
  and a hypothesis concerning its origin}.
\newblock {\em Bull. Astron. Inst. Neth.}{ \bf 11}, 91--110 (1950).

\bibitem{heitre86}
{Heisler}, J. and {Tremaine}, S.
\newblock {The influence of the galactic tidal field on the Oort comet cloud}.
\newblock {\em Icarus}{ \bf 65}, 13--26 (1986).

\bibitem{kaibquinn09}
{Kaib}, N.~A. and {Quinn}, T.
\newblock {Reassessing the Source of Long-Period Comets}.
\newblock {\em Science}{ \bf 325}, 1234--1236 (2009).

\bibitem{adamslaugh01}
{Adams}, F.~C. and {Laughlin}, G.
\newblock {Constraints on the Birth Aggregate of the Solar System}.
\newblock {\em Icarus}{ \bf 150}, 151--162 (2001).

\bibitem{zaktre04}
{Zakamska}, N.~L. and {Tremaine}, S.
\newblock {Excitation and Propagation of Eccentricity Disturbances in Planetary
  Systems}.
\newblock {\em \aj}{ \bf 128}, 869--877 (2004).

\bibitem{cham02}
{Chambers}, J.~E., {Quintana}, E.~V., {Duncan}, M.~J., and {Lissauer}, J.~J.
\newblock {Symplectic Integrator Algorithms for Modeling Planetary Accretion in
  Binary Star Systems}.
\newblock {\em Astron. J.}{ \bf 123}, 2884--2894 (2002).

\bibitem{liss93}
{Lissauer}, J.~J.
\newblock {Planet formation}.
\newblock {\em Ann. Rev. Astron. Astrophys.}{ \bf 31}, 129--174 (1993).

\bibitem{wright09}
{Wright}, J.~T., {Upadhyay}, S., {Marcy}, G.~W., {Fischer}, D.~A., {Ford},
  E.~B., and {Johnson}, J.~A.
\newblock {Ten New and Updated Multiplanet Systems and a Survey of Exoplanetary
  Systems}.
\newblock {\em Astrophys. J.}{ \bf 693}, 1084--1099 (2009).

\bibitem{jurtre08}
{Juri{\'c}}, M. and {Tremaine}, S.
\newblock {Dynamical Origin of Extrasolar Planet Eccentricity Distribution}.
\newblock {\em Astrophys. J.}{ \bf 686}, 603--620 (2008).

\bibitem{fordras08}
{Ford}, E.~B. and {Rasio}, F.~A.
\newblock {Origins of Eccentric Extrasolar Planets: Testing the Planet-Planet
  Scattering Model}.
\newblock {\em Astrophys. J.}{ \bf 686}, 621--636 (2008).

\bibitem{malmdav09}
{Malmberg}, D. and {Davies}, M.~B.
\newblock {On the origin of eccentricities among extrasolar planets}.
\newblock {\em Mon. Not. R. Astron. Soc.}{ \bf 394}, L26--L30 (2009).

\bibitem{ray10}
{Raymond}, S.~N., {Armitage}, P.~J., and {Gorelick}, N.
\newblock {Planet-Planet Scattering in Planetesimal Disks. II. Predictions for
  Outer Extrasolar Planetary Systems}.
\newblock {\em Astrophys. J.}{ \bf 711}, 772--795 (2010).

\bibitem{koz62}
{Kozai}, Y.
\newblock {Secular perturbations of asteroids with high inclination and
  eccentricity}.
\newblock {\em Astron. J.}{ \bf 67}, 591--598 (1962).

\bibitem{hol97}
{Holman}, M., {Touma}, J., and {Tremaine}, S.
\newblock {Chaotic variations in the eccentricity of the planet orbiting 16
  Cygni B}.
\newblock {\em Nature}{ \bf 386}, 254--256 (1997).

\bibitem{wumur03}
{Wu}, Y. and {Murray}, N.
\newblock {Planet Migration and Binary Companions: The Case of HD 80606b}.
\newblock {\em Astrophys. J.}{ \bf 589}, 605--614 (2003).

\bibitem{fabtre07}
{Fabrycky}, D. and {Tremaine}, S.
\newblock {Shrinking Binary and Planetary Orbits by Kozai Cycles with Tidal
  Friction}.
\newblock {\em Astrophys. J.}{ \bf 669}, 1298--1315 (2007).

\bibitem{mar10}
{Marois}, C., {Zuckerman}, B., {Konopacky}, Q.~M., {Macintosh}, B., and
  {Barman}, T.
\newblock {Images of a fourth planet orbiting HR 8799}.
\newblock {\em Nature}{ \bf 468}, 1080--1083 (2010).

\bibitem{sum11}
{Sumi}, T., {Kamiya}, K., {Bennett}, D.~P., {Bond}, I.~A., {Abe}, F.,
  {Botzler}, C.~S., {Fukui}, A., {Furusawa}, K., {Hearnshaw}, J.~B., {Itow},
  Y., {Kilmartin}, P.~M., {Korpela}, A., {Lin}, W., {Ling}, C.~H., {Masuda},
  K., {Matsubara}, Y., {Miyake}, N., {Motomura}, M., {Muraki}, Y., {Nagaya},
  M., {Nakamura}, S., {Ohnishi}, K., {Okumura}, T., {Perrott}, Y.~C.,
  {Rattenbury}, N., {Saito}, T., {Sako}, T., {Sullivan}, D.~J., {Sweatman},
  W.~L., {Tristram}, P.~J., {Udalski}, A., {Szyma{\'n}ski}, M.~K., {Kubiak},
  M., {Pietrzy{\'n}ski}, G., {Poleski}, R., {Soszy{\'n}ski}, I., {Wyrzykowski},
  {\L}., {Ulaczyk}, K., and {Microlensing Observations in Astrophysics (MOA)
  Collaboration}.
\newblock {Unbound or distant planetary mass population detected by
  gravitational microlensing}.
\newblock {\em Nature}{ \bf 473}, 349--352 (2011).

\bibitem{egg07}
{Eggenberger}, A., {Udry}, S., {Chauvin}, G., {Beuzit}, J.-L., {Lagrange},
  A.-M., {S{\'e}gransan}, D., and {Mayor}, M.
\newblock {The impact of stellar duplicity on planet occurrence and properties.
  I. Observational results of a VLT/NACO search for stellar companions to 130
  nearby stars with and without planets}.
\newblock {\em Astron. Astrophys.}{ \bf 474}, 273--291 (2007).

\bibitem{desbar07}
{Desidera}, S. and {Barbieri}, M.
\newblock {Properties of planets in binary systems. The role of binary
  separation}.
\newblock {\em Astron. Astrophys.}{ \bf 462}, 345--353 (2007).

\bibitem{rag06}
{Raghavan}, D., {Henry}, T.~J., {Mason}, B.~D., {Subasavage}, J.~P., {Jao},
  W.-C., {Beaulieu}, T.~D., and {Hambly}, N.~C.
\newblock {Two Suns in The Sky: Stellar Multiplicity in Exoplanet Systems}.
\newblock {\em Astrophys. J.}{ \bf 646}, 523--542 (2006).

\bibitem{mug06}
{Mugrauer}, M., {Neuh{\"a}user}, R., {Mazeh}, T., {Guenther}, E.,
  {Fern{\'a}ndez}, M., and {Broeg}, C.
\newblock {A search for wide visual companions of exoplanet host stars: The
  Calar Alto Survey}.
\newblock {\em Astronomische Nachrichten}{ \bf 327}, 321--327 (2006).

\bibitem{jones06}
{Jones}, H.~R.~A., {Butler}, R.~P., {Tinney}, C.~G., {Marcy}, G.~W., {Carter},
  B.~D., {Penny}, A.~J., {McCarthy}, C., and {Bailey}, J.
\newblock {High-eccentricity planets from the Anglo-Australian Planet Search}.
\newblock {\em Mon. Not. R. Astron. Soc.}{ \bf 369}, 249--256 (2006).

\bibitem{naef01}
{Naef}, D., {Latham}, D.~W., {Mayor}, M., {Mazeh}, T., {Beuzit}, J.~L.,
  {Drukier}, G.~A., {Perrier-Bellet}, C., {Queloz}, D., {Sivan}, J.~P.,
  {Torres}, G., {Udry}, S., and {Zucker}, S.
\newblock {HD 80606 b, a planet on an extremely elongated orbit}.
\newblock {\em Astron. Astrophys.}{ \bf 375}, L27--L30 (2001).

\bibitem{inn97}
{Innanen}, K.~A., {Zheng}, J.~Q., {Mikkola}, S., and {Valtonen}, M.~J.
\newblock {The Kozai Mechanism and the Stability of Planetary Orbits in Binary
  Star Systems}.
\newblock {\em Astron. J.}{ \bf 113}, 1915--1919 (1997).

\bibitem{bat11}
{Batygin}, K., {Morbidelli}, A., and {Tsiganis}, K.
\newblock {Formation and evolution of planetary systems in presence of highly
  inclined stellar perturbers}.
\newblock {\em Astron. Astrophys.}{ \bf 533}, A7 (2011).

\bibitem{kaib11}
{Kaib}, N.~A., {Raymond}, S.~N., and {Duncan}, M.~J.
\newblock {55 Cancri: A Coplanar Planetary System That is Likely Misaligned
  with Its Star}.
\newblock {\em Astrophys. J.}{ \bf 742}, L24 (2011).

\end{thebibliography}
\end{document}